\documentstyle[12pt]{article}

\begin{document}

\thispagestyle{empty}
\renewcommand{\thefootnote}{\fnsymbol{footnote}}

\begin{flushright}
{\small
SLAC--PUB--8210\\
July 1999\\}
\end{flushright}

\vspace{1.0cm}

\begin{center}
{\Large\bf 
Nonlinear Longitudinal Waves in High Energy Stored Beams
\footnote{Work supported by
Department of Energy contract  DE--AC03--76SF00515.}}

\vspace{1.0cm}

Stephan I. Tzenov\\
{\it
Stanford Linear Accelerator Center, Stanford University,
Stanford, CA  94309\\}

\end{center}

\vfill

\begin{center}
{\bf\large   
Abstract }
\end{center}

\begin{quote}
We solve the Vlasov equation for the longitudinal 
distribution function and find stationary wave patterns 
when the distribution in the energy error is Maxwellian. 
In the long wavelength limit a stability criterion for 
linear waves has been obtained and a Korteweg-de Vries-
Burgers equation for the relevant hydrodynamic quantities 
has been derived.
\end{quote}

\vfill

\begin{center} 
{\it Paper presented at} 
{\it Workshop on Instabilities of High Intensity Hadron 
Beams in Rings,} \\
{\it Brookhaven National Laboratory} \\
{\it June 28 -- July 1, 1999.} \\

\end{center}

\newpage

\section{Introduction.}

Nonlinear wave interaction in high energy synchrotrons has 
recently received a great deal of attention (see e.g. 
\cite{colsptz}, \cite{tzcol}, \cite{tzenov}), since it has 
proven its importance for understanding a variety of phenomena 
in high intensity beams.

Perhaps, the simplest problem to study is the evolution in 
longitudinal direction only of a intense coasting beam 
influenced by a broad-band resonator type impedance. This model 
exhibits a surprisingly vast variety of interesting features, 
part of which have already been experimentally observed and 
theoretically investigated \cite{colsptz}, \cite{tzcol}, 
\cite{tzenov}. Different types of beam equilibria can be detected 
due to the collective (nonlinear) interaction between beam 
particles and resonator waves, the latter being induced by the 
beam itself. Solutions describing similar types of plasma equilibria 
[Bernstein-Greene-Kruskal (BGK) modes] are well-known in plasma 
physics \cite{bgk}. Structures of arbitrary shape can be formed 
in the nonlinear stationary regime, which substantially depend 
on the type of the initial velocity distribution.

It is the purpose of the present paper to apply techniques 
borrowed from plasma physics to study nonlinear patterns in 
coasting beams that are in close analogy with BGK modes. In 
sections 3 and 4 we solve the Vlasov equation by expanding the 
distribution function in a power series of the resonator potential 
\cite{karimov}, and in the case of initial Maxwellian energy error 
distribution we obtain an equation, describing the evolution of 
stationary waves on the resonator. In section 5 we find a stability 
criterion for linear waves in the long wavelength limit and derive 
a Korteweg-de Vries-Burgers equation for the beam density, current 
velocity and resonator voltage.

\section{Model.}

We consider the longitudinal dynamics of a high energy stored 
beam governed by the set of equations \cite{tzcol}, \cite{tzenov}:

$$
\frac{\partial f}{\partial T } + v \frac{\partial f}{\partial \theta } 
+ \lambda V \frac{\partial f}{\partial v } = 0,
\eqno (2.1)
$$

$$
{\frac{{\partial^2} V}{\partial {T^2}}} + 2 \gamma {\frac{\partial V}
{\partial T }} + {\omega^2} V = {\frac{\partial I}{\partial T}},
\eqno (2.2)
$$

$$
I{\left( \theta; T \right)} = \int dv v 
f{\left( \theta, v; T \right)}. 
\eqno (2.3)
$$

\noindent
The first equation (2.1) is the Vlasov equation for the longitudinal 
distribution function $f{\left( \theta, v; T \right)}$ of an unbunched 
beam, while the second equation (2.2) governs the variation per turn 
of the voltage $V{\left( \theta; T \right)}$ on a resonator. All 
dependent and independent variables, as well as free parameters in 
equations (2.1-3) are dimensionless and have been rescaled according 
to the relations:

$$
T = {\omega_s}t \qquad ; \qquad 
v = {\frac 1 {\omega_s}}{\frac {d \theta} {d t}} = 
1 + {\frac {k_0 \Delta E} {\omega_s}} \qquad ; \qquad
\omega = {\frac {\omega_R} {\omega_s}},
\eqno (2.4a)
$$

$$
\gamma = {\frac \omega {2Q}} \qquad ; \qquad
\lambda = {\frac {e^2 {\cal R} \gamma k_0 \rho_0} \pi}.
\eqno (2.4b)
$$

\noindent
Here $\omega_s$ is the angular revolution frequency of the synchronous 
particle, $\Delta E$ is the energy error, $\omega_R$ is the resonant 
frequency, $Q$ is the quality factor of the resonator, ${\cal R}$ is 
the resonator shunt impedance and $\rho_0$ is the uniform beam density 
distribution in the thermodynamic limit. Furthermore

$$
k_0 = -{\frac {\eta \omega_s} {\beta_s^2 E_s}}
\eqno (2.5)
$$

\noindent
is the proportionality constant between the frequency deviation of a 
non synchronous particle with respect to the synchronous one, while 
$\eta = \alpha_M - \gamma_s^{-2}$ ($\alpha_M$ - momentum compaction 
factor) is the phase slip coefficient. The voltage variation per turn 
$V{\left( \theta; T \right)}$, the beam current 
$I{\left( \theta; T \right)}$ and the longitudinal distribution 
function $f{\left( \theta, v; T \right)}$ entering equations (2.1-3) 
have been rescaled as well from their actual values 
$V_a{\left( \theta; T \right)}$, $I_a{\left( \theta; T \right)}$ and 
$f_a{\left( \theta, v; T \right)}$ as follows:

$$
V_a = 2e \omega_s \rho_0 \gamma {\cal R} V \qquad ; \qquad
I_a = e \omega_s \rho_0 I \qquad ; \qquad
f_a = \rho_0 f.
\eqno (2.6)
$$

\noindent
From the Vlasov equation (2.1) it is straightforward to obtain the 
continuity equation:

$$
{\frac{\partial }{\partial T }} \int dv f + 
{\frac{\partial }{\partial \theta }} \int dvv f = 0, 
\eqno (2.7)
$$

\noindent
which will be needed for the exposition in the next section.

\section{Solution of the Vlasov Equation.}

Let us now try to solve the Vlasov equation by the simple separation 
of variables ansatz:

$$
f{\left( \theta, v; T \right)} = g{\left( v \right)} 
\psi{\left( \theta; T \right)}.
\eqno (3.1)
$$

\noindent
Substitution of (3.1) into the continuity equation (2.7) yields:

$$
{\frac{\partial \psi }{\partial T }} + 
\Omega {\frac{\partial \psi }{\partial \theta }} = 0, 
\eqno (3.2)
$$

\noindent
where

$$
\Omega = {\frac{\int dvv g{\left( v \right)} }
{\int dv g{\left( v \right)} }}. 
\eqno (3.3)
$$

\noindent
The Vlasov equation (2.1) with (3.1-3) in hand can be further 
transformed to

$$
{\frac{\partial \psi }{\partial \theta }} = 
{\frac {\lambda V \psi} {g {\left( \Omega - v \right)} }}
{\frac {dg} {dv}}.
$$

\noindent
The separation of variables ansatz (3.1) implies

$$
{\frac {dg} {dv}} = 
{\frac {\Omega - v} {\sigma_v^2}} g, 
\eqno (3.4a)
$$

\noindent
which leads to the well-known equilibrium Maxwell-Boltzmann 
distribution:

$$
g{\left( v \right)} = {\frac 1 {{\sigma_v} {\sqrt{2 \pi}}}} 
\exp {\left[ - {\frac {{\left( v - \Omega \right)}^2} 
{2 \sigma_v^2}} \right]},
\eqno (3.4)
$$

$$
\psi{\left( \theta; T \right)} = {\cal Z} 
\exp {\left[ {\frac { \lambda \varphi {\left( \theta; T\right)}} 
{\sigma_v^2}} \right]} \qquad ; \qquad
V{\left( \theta; T \right)} = 
{\frac {\partial \varphi{\left( \theta; T \right)}} 
{\partial \theta}},
\eqno (3.5)
$$

\noindent
where

$$
{{\cal Z}^{-1}} = {\int\limits_0^{2 \pi}} d \theta 
\exp {\left[ {\frac { \lambda \varphi {\left( \theta; T\right)}} 
{\sigma_v^2}} \right]}.
\eqno (3.6)
$$

The solution (3.4-6) suggests further generalization \cite{karimov} 
of the separation of variables ansatz (3.1)

$$
f{\left( \theta, v; T \right)} = {\sum\limits_{k=0}^{\infty}} 
g_k{\left( v \right)} 
\varphi^k{\left( \theta; T \right)}.
\eqno (3.7)
$$

\noindent
Instead of equations (3.2) and (3.3) we now have

$$
{\frac{\partial \varphi }{\partial T }} + 
\Omega {\left( \theta; T \right)} 
{\frac{\partial \varphi }{\partial \theta }} = 0, 
\eqno (3.2a)
$$

\noindent
where

$$
\Omega{\left( \theta; T \right)} = 
{\frac{\sum\limits_{k=1}^{\infty} k {{\cal A}_k} 
\varphi^{k-1}{\left( \theta; T \right)} }
{\sum\limits_{k=1}^{\infty} k {{\cal B}_k} 
\varphi^{k-1}{\left( \theta; T \right)} }}. 
\eqno (3.3a)
$$

$$
{\cal A}_k = {\int dvv g_k{\left( v \right)}} 
\qquad ; \qquad 
{\cal B}_k = {\int dv g_k{\left( v \right)}}. 
\eqno (3.8)
$$

\noindent
In order to determine the yet unknown functions 
$g_k{\left( v \right)}$ we make the assumption:

$$
\Omega{\left( \theta; T \right)} = const,
\eqno (3.9)
$$

\noindent
which will be proved {\it a posteriori} to hold and substitute 
(3.7) into the Vlasov equation (2.1). Taking into account (3.2a) 
we obtain:

$$
{\left( v - \Omega \right)} {\sum\limits_{k=1}^{\infty}} 
k g_k{\left( v \right)} 
\varphi^{k-1}{\left( \theta; T \right)} + 
\lambda {\sum\limits_{k=0}^{\infty}} 
{\frac {dg_k{\left( v \right)}} {dv}} 
\varphi^k{\left( \theta; T \right)} = 0. 
\eqno (3.10)
$$

\noindent
Equating coefficients in front of powers of $\varphi$ yields the 
following recurrence relation

$$
{\left( v - \Omega \right)} {\left( k+1 \right)} 
g_{k+1}{\left( v \right)} = 
- \lambda {\frac {dg_k{\left( v \right)}} {dv}}, 
$$

\noindent
or

$$ 
g_{k+1}{\left( v \right)} = 
{\frac {\lambda} {k+1}} {\widehat{\cal D}}
g_k{\left( v \right)}, 
\eqno (3.11)
$$

\noindent
where we have introduced the operator \cite{karimov}

$$
{\widehat{\cal D}} = 
{\frac 1 {\Omega - v}}{\frac d {dv}}. 
\eqno (3.12)
$$

\noindent
Noting that the formal solution of the recurrence relation 
(3.11) has the form

$$
g_k{\left( v \right)} = 
{\frac {\lambda^k} {k!}} {\widehat{\cal D}}^k 
g_0{\left( v \right)} 
\eqno (3.13)
$$

\noindent
we finally arrive at the general solution of the Vlasov 
equation

$$
f{\left( \theta, v; T \right)} = {\sum\limits_{k=0}^{\infty} 
{\frac {\lambda^k {\varphi^k{\left( \theta; T \right)}}} 
{k!}} 
{\widehat{\cal D}}^k
g_0{\left( v \right)}}. 
\eqno (3.14)
$$

\noindent
What remains now is to verify the condition (3.9). It suffices 
to note that \cite{karimov}

$$
{\cal A}_k = {\frac {\lambda^k} {k!}} 
\int dvv {\widehat{\cal D}}^k g_0{\left( v \right)} = 
{\frac {\lambda^k} {k!}}
\int dv {\frac v {\Omega - v}} {\frac d {dv}} 
{\left[ {\widehat{\cal D}}^{k-1} g_0{\left( v \right)} 
\right]} = 
$$

$$
= -{\frac {\lambda^k \Omega} {k!}}
\int {\frac {dv} {{\left( \Omega - v \right)}^2}} 
{{\widehat{\cal D}}^{k-1} g_0{\left( v \right)}}, 
$$

\noindent
and similarly

$$
{\cal B}_k = 
-{\frac {\lambda^k} {k!}}
\int {\frac {dv} {{\left( \Omega - v \right)}^2}} 
{{\widehat{\cal D}}^{k-1} g_0{\left( v \right)}}.
$$

\noindent
Thus

$$
{\cal A}_k = \Omega {\cal B}_k, 
\eqno (3.15)
$$

\noindent
which proves equation (3.9).

Clearly the solution (3.14) is uniquely determined by the 
generic function $g_0{\left( v \right)}$. The simplest choice 
is when $g_0{\left( v \right)}$ is the Maxwellian (3.4), that 
is $g_0{\left( v \right)}$ itself is an eigenfunction of the 
operator ${\widehat{\cal D}}$ with an eigenvalue 
${\sigma_v^{-2}}$ [c.f. equation (3.4a)]. In this case we 
immediately recover the distribution (3.1) with (3.4-6).

\section{Nonlinear Stationary Waves.}

In order to derive an equation for the potential 
$\varphi{\left( \theta; T \right)}$ we insert (3.1) and (3.4-6) 
into (2.2) and obtain:

$$
{\frac {\partial^3 \varphi} {\partial \theta \partial T^2}} + 
2 \gamma 
{\frac {\partial^2 \varphi} {\partial \theta \partial T}} +
\omega^2 {\frac {\partial \varphi} {\partial \theta}} = 
{\cal Z} \Omega {\frac {\partial} {\partial T}} {\left[
\exp {\left( {\frac {\lambda \varphi} {\sigma_v^2}} \right)}
\right]}. 
\eqno (4.1)
$$

\noindent
Making use of relation (3.2a) we cast equation (4.1) into the 
form

$$
{\frac {\partial^3 \varphi} {\partial T^3}} + 
2 \gamma 
{\frac {\partial^2 \varphi} {\partial T^2}} +
\omega^2 {\frac {\partial \varphi} {\partial T}} = 
- {\cal Z} {\Omega^2} {\frac {\partial} {\partial T}} 
{\left[
\exp {\left( {\frac {\lambda \varphi} {\sigma_v^2}} \right)}
\right]}. 
\eqno (4.2)
$$

\noindent
Integrating once equation (4.2) with due account of the initial 
condition

$$
\varphi{\left( \theta; T= 0 \right)} = 
{\frac {\partial \varphi{\left( \theta; T= 0 \right)}} 
{\partial T}} = 
{\frac {\partial^2 \varphi{\left( \theta; T= 0 \right)}} 
{\partial T^2}} = 0
\eqno (4.3)
$$

\noindent
we obtain

$$
{\frac {\partial^2 \varphi} {\partial T^2}} + 
2 \gamma 
{\frac {\partial \varphi} {\partial T}} +
\omega^2 {\varphi} = 
{\cal Z} {\Omega^2} 
{\left[ 1 - 
\exp {\left( {\frac {\lambda \varphi} {\sigma_v^2}} \right)}
\right]}. 
\eqno (4.4)
$$

\noindent
Expanding the factor in square brackets on the right-hand-side 
of equation (4.4) around the stationary solution $\varphi_s = 0$ 
yields

$$
{\frac {\partial^2 \varphi} {\partial T^2}} + 
2 \gamma 
{\frac {\partial \varphi} {\partial T}} +
\omega^2 {\varphi} = 
- {\frac {\lambda {\cal Z} {\Omega^2}} {\sigma_v^2}} 
\varphi {\left( 1 + 
{\frac {\lambda \varphi} {2 \sigma_v^2}} + 
{\frac {\lambda^2 \varphi^2} {6 \sigma_v^4}} + 
...
\right)}. 
\eqno (4.5)
$$

Above the transition energy $\gamma_s > \gamma_T$ 
${\left( \gamma_T = \alpha_M^{-1/2}  \right)}$  the parameter 
$\lambda$ is negative, so that two cases can be distinguished. 
Defining

$$
\omega_0 = \omega^2 - 
\frac {|\lambda|{\cal Z} \Omega^2} {\sigma_v^2}, 
\eqno (4.6)
$$

\noindent
we can state the two cases mentioned above in a more explicit 
way:

Case {\bf I}: Provided $\omega_0 > 0$, equation (4.5) can be 
transformed to a damped Duffing equation with an additional 
quadratic nonlinearity

$$
{\frac {\partial^2 \varphi} {\partial T^2}} + 
2 \gamma 
{\frac {\partial \varphi} {\partial T}} + 
{\left| \omega_0 \right|} {\varphi} = 
- {\frac {\lambda^2 {\cal Z} {\Omega^2}} {2 \sigma_v^4}} 
{\left( \varphi^2 - 
{\frac {|\lambda|} {3 \sigma_v^2}} \varphi^3 
\right)}. 
\eqno (4.7)
$$

Case {\bf II}: For $\omega_0 < 0$ equation (4.5) takes the form

$$
{\frac {\partial^2 \varphi} {\partial T^2}} + 
2 \gamma 
{\frac {\partial \varphi} {\partial T}} - 
{\left| \omega_0 \right|} {\varphi} = 
- {\frac {\lambda^2 {\cal Z} {\Omega^2}} {2 \sigma_v^4}} 
{\left( \varphi^2 - 
{\frac {|\lambda|} {3 \sigma_v^2}} \varphi^3 
\right)}. 
\eqno (4.8)
$$

\noindent
In the limit $\gamma \rightarrow 0$ equation (4.8) can be 
solved when neglecting the cubic term. The result is:

$$
\varphi{\left( \theta; T \right)} = 
{\frac {3 |\omega_0| \sigma_v^4} 
{\lambda^2 {\cal Z} \Omega^2 \cosh^2 {\left[ 
{\frac {\sqrt{|\omega_0|}} {2 \Omega}}
{\left( \theta - \Omega T \right)}
\right]}}}.
\eqno (4.9)
$$

\noindent
This is a drifting hump-like structure that is well-known 
as a solitary wave of the Korteweg-de Vries (KdV) type.

\section{The Korteweg-de Vries-Burgers Equation.}

The exact solution of the Vlasov equation obtained in the 
preceding sections was found based on the stationary wave 
condition given by the continuity equation (3.2). In order 
to provide a more general treatment of the problem we 
introduce the new coordinates and variables along with the 
moving beam particles

$$
z = \theta - T  \qquad ; \qquad  u = v - 1.
\eqno (5.1)
$$

\noindent
Then the basic equations (2.1-3) can be written as:

$$
\frac{\partial f}{\partial \theta } + u \frac{\partial f}
{\partial z } + \lambda V \frac{\partial f}{\partial u } = 0,
\eqno (5.2)
$$

$$
{\frac{{\partial^2} V}{\partial {z^2}}} - 2 \gamma {\frac{\partial V}
{\partial z }} + {\omega^2} V = - {\frac{\partial}{\partial z}} 
{\int du {\left( 1 + u \right)} f{\left( z, u; \theta \right)}}.
\eqno (5.3)
$$

\noindent
Let us now pass to the hydrodynamic description of the 
longitudinal beam motion. The gas dynamic equations read as

$$
{\frac {\partial F} {\partial \theta}} + {\frac {\partial} 
{\partial z}} {\left( FU  \right)} = 0, 
\eqno (5.4)
$$

$$
{\frac {\partial U} {\partial \theta}} + U {\frac {\partial U} 
{\partial z}} = \lambda V - {\frac {\sigma_v^2} F} 
{\frac {\partial F} {\partial z}}, 
\eqno (5.5)
$$

$$
{\frac{{\partial^2} V}{\partial {z^2}}} - 2 \gamma {\frac{\partial V}
{\partial z }} + {\omega^2} V = {\frac{\partial F}{\partial \theta}} - 
{\frac {\partial F} {\partial z}}, 
\eqno (5.6)
$$

\noindent
where

$$
F{\left( z; \theta \right)} = 
\int du f{\left( z, u; \theta \right)} \qquad ; \qquad
F{\left( z; \theta \right)} U{\left( z; \theta \right)} = 
\int du u f{\left( z, u; \theta \right)}.
\eqno (5.7)
$$

\noindent
Obviously the stationary solution of the gas dynamic equations 
(5.4-6) is given by

$$
F_0 = 1  \qquad ; \qquad  U_0 = 0  \qquad ; \qquad 
V_0 = 0. 
$$

\noindent
The dispersion law of linear waves of the form

$$
{\left( F, U, V \right)} = 
{\left( F_L, U_L, V_L \right)} 
\exp {\left[ i \left( \Omega \theta - k z \right) 
\right]}
$$

\noindent
is governed by the following equation

$$
1 - i \lambda Z{\left( k \right)} 
{\frac {k + \Omega} {\Omega^2 - k^2 \sigma_v^2}} = 0, 
\eqno (5.8)
$$

\noindent
where $Z{\left( k \right)}$ is the well-known impedance 
function

$$
Z{\left( k \right)} = {\frac {i k} 
{k^2 + 2 i \gamma k - \omega^2}}. 
\eqno (5.9)
$$

\noindent
In the long wavelength limit (small $k$) the dispersion 
equation (5.8) has two roots given by the expression

$$
\Omega_{1,2} = {\frac k {2 \omega^2}} 
{\left( \lambda \pm \sqrt{\lambda^2 + 4 \lambda \omega^2 + 
4 \omega^4 \sigma_v^2} \right)}, 
\eqno (5.10)
$$

\noindent
which are real below transition energy. However, the situation 
when the energy of the synchronous particle is above transition 
energy is different. The solutions (5.10) to the dispersion 
equation are real, provided

$$
|\lambda| \leq \lambda_1  \qquad ; \qquad  
|\lambda| \geq \lambda_2  \qquad ; \qquad  
\lambda_{1,2} = 2 \omega^2 {\left( 1 \mp 
\sqrt{1 - \sigma_v^2} \right)}. 
\eqno (5.11)
$$

\noindent
An instability occurs when $\Omega_{1,2}$ are complex, that 
is when

$$
\lambda_1 < |\lambda| < \lambda_2. 
\eqno (5.12)
$$

\noindent
In what follows we will study the case when our system is 
linearly stable, that is either below transition energy or in 
the stability region (5.11).

The solution of the dispersion equation in the long wavelength 
limit suggests that new scaled coordinates should be introduced 
\cite{washimi}, \cite{debnath}

$$
\sigma = \sqrt{\epsilon} 
{\left( z - \alpha \theta \right)}  \qquad ; \qquad  
\chi = \epsilon^{3/2} \theta, 
\eqno (5.13)
$$

\noindent
where $\epsilon$ is a formal small parameter. Then the gas 
dynamic equations can be rewritten as

$$
- \alpha {\frac {\partial F} {\partial \sigma}} + 
{\frac {\partial} {\partial \sigma}} {\left( FU  \right)} + 
\epsilon {\frac {\partial F} {\partial \chi}} = 0, 
\eqno (5.14)
$$

$$
- \alpha {\frac {\partial U} {\partial \sigma}} + 
U {\frac {\partial U} {\partial \sigma}} + 
\epsilon {\frac {\partial U} {\partial \chi}} 
= \lambda \widetilde{V} - {\frac {\sigma_v^2} F} 
{\frac {\partial F} {\partial \sigma}}, 
\eqno (5.15)
$$

$$
\epsilon {\frac{{\partial^2} \widetilde{V}}{\partial {\sigma^2}}} - 
2 \epsilon \gamma_0 {\frac{\partial \widetilde{V}}
{\partial \sigma }} + {\omega^2} \widetilde{V} = 
- {\left( 1 + \alpha \right)}
{\frac{\partial F}{\partial \sigma}} + 
\epsilon {\frac {\partial F} {\partial \chi}}, 
\eqno (5.16)
$$

\noindent
where

$$
V = {\sqrt{\epsilon}} {\widetilde{V}}  \qquad ; \qquad  
\gamma = {\sqrt{\epsilon}} \gamma_0, 
\eqno (5.17)
$$

$$
\omega^2 \alpha^2 - \lambda \alpha - \lambda - 
\omega^2 \sigma_v^2 = 0  \qquad ; \qquad  
{\left( \alpha = \frac {\Omega_{1,2}} {k} \right)}. 
\eqno (5.18)
$$

\noindent
Assuming the perturbation expansions:

$$
F = 1 + {\sum\limits_{m=1}^{\infty}} {\epsilon^m}
F_m  \qquad ; \qquad  
U = {\sum\limits_{m=1}^{\infty}} {\epsilon^m} U_m  
\qquad ; \qquad  
\widetilde{V} = {\sum\limits_{m=1}^{\infty}} {\epsilon^m} 
V_m
\eqno (5.19)
$$

\noindent
for the first and second-order terms in $\epsilon$ we 
obtain respectively

$$
\frac {\partial U_1} {\partial \sigma} = 
\alpha {\frac {\partial F_1} {\partial \sigma}} = 
{\frac {\alpha \lambda V_1} {\sigma_v^2 - \alpha^2}}, 
\eqno (5.20)
$$

\noindent
or

$$
U_1{\left( \sigma, \chi \right)} = 
\alpha F_1{\left( \sigma, \chi \right)} + 
G{\left( \chi \right)}, 
\eqno (5.21)
$$

\noindent
where $G{\left( \chi \right)}$ is a generic function of 
the variable $\chi$, and

$$
- \alpha {\frac {\partial F_2} {\partial \sigma}} + 
{\frac {\partial U_2} {\partial \sigma}} + 
{\frac {\partial} {\partial \sigma}} {\left( F_1 U_1 \right)} + 
{\frac {\partial F_1} {\partial \chi}} = 0, 
\eqno (5.22a)
$$

$$
- \alpha {\frac {\partial U_2} {\partial \sigma}} + 
U_1 {\frac {\partial U_1} {\partial \sigma}} + 
{\frac {\partial U_1} {\partial \chi}} 
= \lambda V_2 - {\sigma_v^2} 
{\frac {\partial F_2} {\partial \sigma}} + 
{\sigma_v^2} F_1 {\frac {\partial F_1} {\partial \sigma}}, 
\eqno (5.22b)
$$

$$
{\frac{{\partial^2} V_1}{\partial {\sigma^2}}} - 
2 \gamma_0 {\frac{\partial V_1}
{\partial \sigma }} + {\omega^2} V_2 = 
- {\left( 1 + \alpha \right)}
{\frac{\partial F_2}{\partial \sigma}} + 
{\frac {\partial F_1} {\partial \chi}}. 
\eqno (5.22c)
$$

\noindent
Eliminating $F_2$, $U_2$ and $V_2$ from equations (5.22) 
we finally arrive at the Korteweg-de Vries-Burgers equation

$$
{\frac {\partial F_1} {\partial \chi}} + 
{\left( c_1 F_1 + c_2 G \right)} 
{\frac {\partial F_1} {\partial \sigma}} + 
D {\frac {\partial^3 F_1} {\partial \sigma^3}} - 
2 \gamma D {\frac {\partial^2 F_1} {\partial \sigma^2}} = 
h {\frac {d G} {d \chi}}, 
\eqno (5.23)
$$

\noindent
where

$$
c_1 = {\frac {\omega^2 {\left( 3 \alpha^2 - \sigma_v^2 \right)}} 
{2 \alpha \omega^2 - \lambda}}  \qquad ; \qquad  
c_2 = {\frac {2 \alpha \omega^2} 
{2 \alpha \omega^2 - \lambda}}, 
\eqno (5.24a)
$$

$$
D = {\frac {\sigma_v^2 - \alpha^2} 
{2 \alpha \omega^2 - \lambda}}  \qquad ; \qquad  
h = {\frac {\omega^2} 
{\lambda - 2 \alpha \omega^2}}. 
\eqno (5.24b)
$$

\noindent
It is important to note that $\alpha^{-1}{U_1}$ and 
$\lambda {\left( \sigma_v^2 - \alpha^2 \right)}^{-1}
{\int d \sigma V_1}$ satisfy exactly the same equation (5.23).

Similar Korteweg-de Vries-Burgers equation in the case below 
transition energy has been recently derived by A. Aceves 
employing the method of multiple scales \cite{aceves}.

\section{Concluding Remarks.}

We have studied the longitudinal dynamics of a high energy 
coasting beam moving in a resonator. The coupled Vlasov equation 
for the longitudinal distribution function and the equation for 
the resonator voltage have been solved by closely following the 
method of Karimov and Lewis \cite{karimov}. The key point of 
this method consists in the representation of the distribution 
function as a power series in the resonator potential. Further 
self-consistent stationary wave patterns have been found in the 
simplest equilibrium case of Maxwellian distribution in the 
energy error.

In the long wavelength (small wavenumber) limit a stability 
criterion for linear waves has been obtained and a Korteweg-de 
Vries-Burgers equation for the relevant hydrodynamic quantities 
has been derived.

An important (and interesting) extension of the results obtained
here involves the longitudinal dynamics of a bunched beam. These 
will be reported elsewhere.

\section{Acknowledgements.}

I would like to thank A. Aceves and P. Colestock for many 
helpful discussions concerning the subject of the present paper. 

This work was supported by the US Department of Energy, Office 
of Basic Energy Sciences, under contract DE-AC03-76SF00515.

\end{document}